\documentclass[aip, jmp, amsmath,amssymb,reprint,]{revtex4-1}

\usepackage{graphicx}
\usepackage{dcolumn}
\usepackage{bm}
\usepackage[version=3]{mhchem} 
\usepackage[T1]{fontenc}       
\usepackage{amsmath}
\usepackage{upgreek}
\usepackage{natbib}
\usepackage{siunitx}

\begin{document}


\title{Self-aligned local electrolyte gating of 2D materials with nanoscale resolution}
\author{Cheng Peng}
\thanks{Contributed equally to this work}
\affiliation{Department of ELectrical Engineering and Computer Science, Massachusetts Institute of Technology, Cambridge, Massachusetts 02139, United States}
\author{Dmitri K. Efetov}
\thanks{Contributed equally to this work}
\affiliation{Department of ELectrical Engineering and Computer Science, Massachusetts Institute of Technology, Cambridge, Massachusetts 02139, United States}
\author{Sebastien Nanot}
\thanks{Contributed equally to this work}
\affiliation{ICFO-Institut de Ciencies Fotoniques, The Barcelona Institute of Science and Technology, 08860 Castelldefels (Barcelona), Spain}
\author{Ren-Jye Shiue}
\author{Gabriele Grosso}
\affiliation{Department of ELectrical Engineering and Computer Science, Massachusetts Institute of Technology, Cambridge, Massachusetts 02139, United States}
\author{Yafang Yang}
\affiliation{Department of Physics, Massachusetts Institute of Technology, Cambridge, Massachusetts 02139, United States}
\author{Marek Hempel}
\affiliation{Department of ELectrical Engineering and Computer Science, Massachusetts Institute of Technology, Cambridge, Massachusetts 02139, United States}
\author{Pablo Jarillo-Herrero}
\affiliation{Department of Physics, Massachusetts Institute of Technology, Cambridge, Massachusetts 02139, United States}
\author{Jing Kong}
\affiliation{Department of ELectrical Engineering and Computer Science, Massachusetts Institute of Technology, Cambridge, Massachusetts 02139, United States}
\author{Frank H. L. Koppens}
\affiliation{ICFO-Institut de Ciencies Fotoniques, The Barcelona Institute of Science and Technology, 08860 Castelldefels (Barcelona), Spain}
\affiliation{ICREA-Instituci\'{o} Catalana de Recer\c{c}a i Estudis Avancats, 08010 Barcelona, Spain.}
\author{Dirk Englund}
\email{englund@mit.edu}
\affiliation{Department of ELectrical Engineering and Computer Science, Massachusetts Institute of Technology, Cambridge, Massachusetts 02139, United States}

\date{\today}

\begin{abstract}
In the effort to make 2D materials-based devices smaller, faster, and more efficient, it is important to control charge carrier at lengths approaching the nanometer scale. Traditional gating techniques based on capacitive coupling through a gate dielectric cannot generate strong and uniform electric fields at this scale due to divergence of the fields in dielectrics. This field divergence limits the gating strength, boundary sharpness, and pitch size of periodic structures, and restricts possible geometries of local gates (due to wire packaging), precluding certain device concepts, such as plasmonics and transformation optics based on metamaterials. Here we present a new gating concept based on a dielectric-free self-aligned electrolyte technique that allows spatially modulating charges with nanometer resolution. We employ a combination of a solid-polymer electrolyte gate and an ion-impenetrable e-beam-defined resist mask to locally create excess charges on top of the gated surface. Electrostatic simulations indicate high carrier density variations of $\Delta n = \SI{e14}{cm^{-2}}$ across a length of \SI{10}{nm} at the mask boundaries on the surface of a 2D conductor, resulting in a sharp depletion region and a strong in-plane electric field of \SI{6d8}{V.m^{-1}} across the so-created junction. We apply this technique to the 2D material graphene to demonstrate the creation of tunable \emph{p-n} junctions for optoelectronic applications. We also demonstrate the spatial versatility and self-aligned properties of this technique by introducing a novel graphene thermopile photodetector.
\end{abstract}

\maketitle

Modulation of charge carrier concentration of semiconductors lies at the heart of many electronic and optoelectronic device operation principles~\cite{neamen2003semiconductor,chuang2012physics}. This modulation is especially essential for two-dimensional (2D) van der Waals materials~\cite{ferrari2015science,bao2012graphene,avouris2014graphene,koppens2014photodetectors,wang2012electronics,britnell2013strong,liu2014graphene,sassi2016graphene} where it is usually much stronger (from 0.15 electrons/cell to 15 electrons/cell) compared to bulk materials and can be dynamically tuned with electrostatic gating methods. In recent years, rapidly developing device concepts and applications impose stronger and stronger requirements on the spatial resolution and highest-achievable carrier concentration of gating techniques. For example, a spatially sharp ($\sim$\SI{10}{nm}) \emph{p-n} junction and a high carrier density contrast across the junction is the key to the realization of concepts such as tunnel diodes~\cite{roy2015dual} and negative electron refractive index~\cite{cheianov2007focusing}. A strong in-plane electric field across the junction as a result of the junction sharpness facilitates electron-hole pair separation in the photovoltaic (PV) effect\cite{shang2011effect} and thus can improve the quantum efficiency of PV-based solar cells and photodetectors~\cite{konstantatos2010nanostructured}. Many novel device concepts also rely on the ability to create metamaterials with spatial carrier density variations down to the nanometer scale, including for instance graphene with periodically doped nanodisk or nanoribbon arrays for complete optical absorption in the visible and near-infrared~\cite{thongrattanasiri2012complete,ye2015electrically}, graphene with doped waveguide, bend and resonator patterns for a plasmon-based nanophotonic network~\cite{vakil2011transformation}, and superlattices based on graphene and other 2D materials for concepts such as electron beam supercollimation~\cite{yankowitz2012emergence,kang2013electronic,park2008anisotropic,park2008electron}. Implementing these concepts calls for a gating method that allows for sharp \emph{p-n} junctions with narrow depletion regions ($\sim$\SI{10}{nm}), large carrier density contrasts ($\SI{e14}{cm^{-2}}$), strong in-plane electric fields (\SI{6d8}{V.m^{-1}}), and the versatility to generate complex spatial doping profiles with a nanoscale resolution.

The state-of-the-art electrostatic gating technique for modulating charge carrier concentration and creating \emph{p-n} junctions is the metal-dielectric split gate technique~\cite{baugher2014optoelectronic,pospischil2014solar,ross2014electrically}. This method is based on the electric field effect~\cite{novoselov2004electric} in which electric voltages are applied across a gate dielectric to induce extra charges on the 2D material surface. A \emph{p-n} junction can be created by applying opposing electric potentials to the two sides of a boundary to induce charges with opposite polarities. Although this technique is convenient, several limitations restrict its use when more extreme requirements are desirable: In terms of carrier density contrast, dielectric-based gating can only induce a carrier concentration variation $\Delta n$ of less than $2\times\SI{e13}{cm^{-2}}$ for typical dielectrics such as \ce{SiO2}, \ce{HfO2}, \ce{SiN}, and hexagonal-\ce{BN}, due to a maximal applicable voltage across the dielectrics before the dielectric breakdown (molecular bond breakage and defects)~\cite{mcpherson2003thermochemical,sire2007statistics,hattori2014layer}. In terms of junction sharpness, the carrier density has a slowly varying profile across the junction due to electric field divergence in dielectrics, with a characteristic length similar to the thickness of the dielectric, making it hard to create sharp junctions at nanoscale unless with extremely thin (a few nanometers) dielectrics which typically has undesirable leakage and tunneling currents. Furthermore, due to wire-packaging difficulties and fabrication limitation of the electrodes, complex gating patterns and device geometries with large numbers of gating electrodes at the nanoscale is practically challenging.

In this paper we present a self-aligned gating concept with a spatial resolution down to sub-\SI{10}{nm} based on electrolytic gating. In contrast to dielectric-based gating, electrolyte gating can concentrate excess charges directly on the surface of the 2D material and reach a capacitance of $C=\SI{3.2}{\micro F.cm^{-2}}$ (250 times higher than a typical \SI{300}{nm} \ce{SiO2} gate)~\cite{efetov2010controlling,ueno2011discovery,ohno2009electrolyte}, which enables carrier density modification up to $\Delta n = \SI{e14}{cm^{-2}}$. Since patterning of electrolyte is challenging, a general method for local electrolyte gates at nanoscale has not been demonstrated so far. To achieve this goal we introduce a lithographical masking technique based on e-beam over-exposed Poly(methyl methacrylate) (PMMA) that can screen ions in electrolyte. This e-beam patterned mask can prevent the mask-protected areas from being in contact with, and thus modulated by, the electrolyte gate. The mask hence effectively creates lithographically-defined local electrolyte gates with versatile geometries and feature sizes down to several nanometers.

\begin{figure*}[!htb]
\centering
\includegraphics[width=0.95\textwidth]{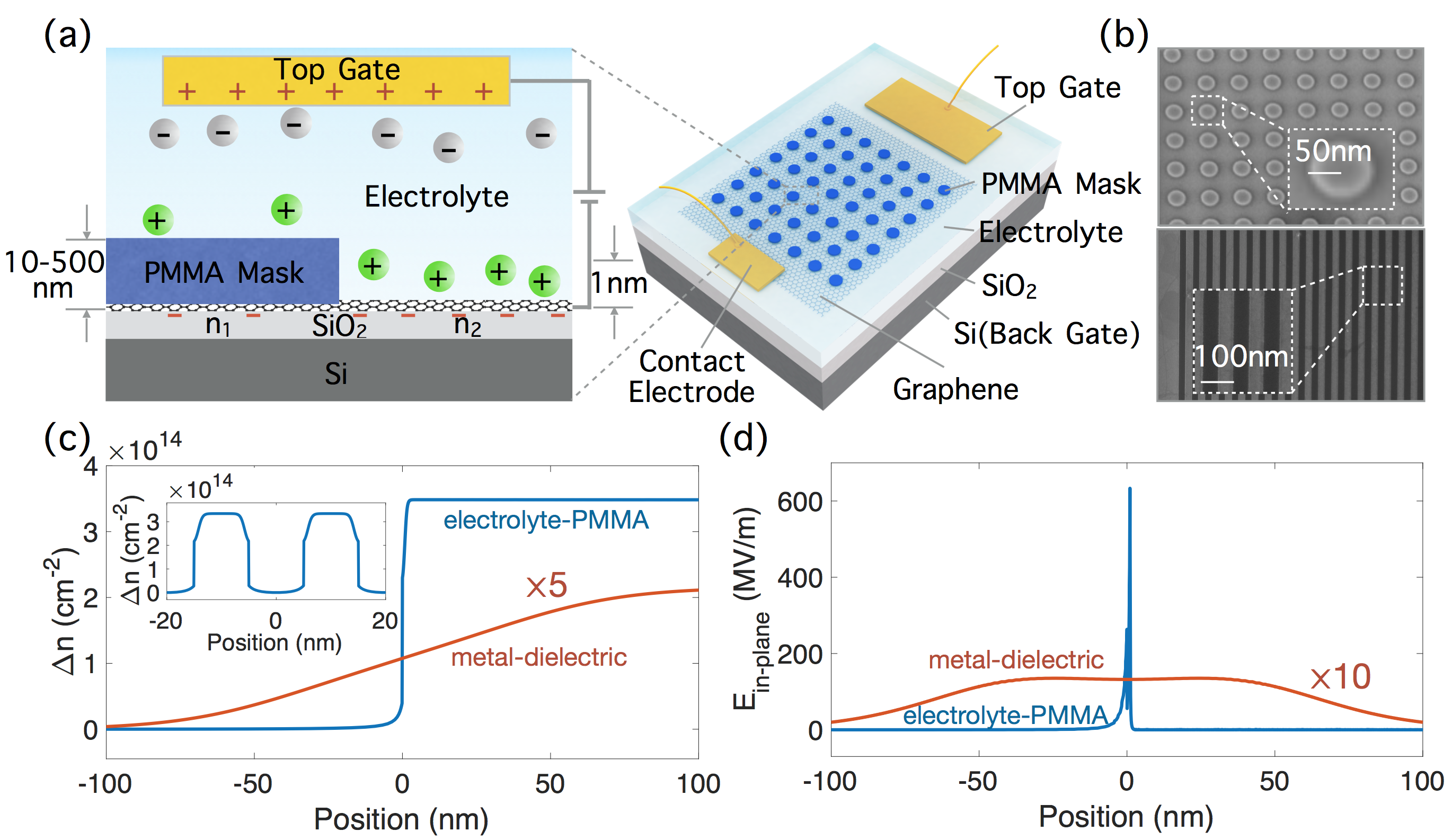}
\caption{\label{schematic-simulation} Self-aligned nanoscale electrolyte gating. (a) Geometry and working principle of nanoscale electrolytic doping of 2D materials with a PMMA screening mask. $n_1$ and $n_2$ denote charge carrier densities in the mask-protected region and the exposed region, respectively. The charge carrier density contrast is defined as $\Delta n = |n_1 - n_2|$. (b) Scanning electron micrographs (SEMs) of example fabricated PMMA masks on graphene with nanoscale dimensions. (c) Simulated charge carrier density $n$ profile for a single junction, compared between the proposed electrolyte-PMMA-mask gating scheme and a metal-dielectric split gating scheme. Inset: Simulated charge carrier density $n$ profile for the proposed electrolyte-PMMA gating scheme with a periodic PMMA mask with width $l=\SI{10}{nm}$ and thickness $d=\SI{10}{nm}$. (d) Simulated in-plane electric field intensity $\si{E_{in-plane}}$ for a single junction, compared between the proposed electrolyte-PMMA-mask gating scheme and a metal-dielectric split gating scheme. Simulation parameters used in (c) and (d) are summarized below: Gate voltages are $\si{V_{tg}} = \SI{1}{V}$ and $\si{V_{bg}} = \SI{-60}{V}$ for the electrolyte-PMMA gating scheme, and $\si{V_{lg}} = \SI{-60}{V}$ and $\si{V_{rg}} = \SI{60}{V}$ for the metal-dielectric gating scheme. The dielectric gate has a thickness of $d=\SI{60}{nm}$. The PMMA mask has a thickness of $d=\SI{100}{nm}$. A gap of \SI{100}{nm} is assumed between the two metal split gates. Additional details on parameters used are in the Supplementary Information.}
\end{figure*}

Figure~\ref{schematic-simulation}(a) illustrates the technique for a graphene sheet. When a voltage is applied between the electrolyte top gate and the graphene, ions in the electrolyte accumulate on the graphene surface, only in regions that are uncovered by the PMMA mask, creating a self-aligned electrolyte gating pattern defined by the shape of the mask. An additional \ce{SiO2} back gate allows weak \emph{p} or \emph{n} doping of the regions covered by the mask.

For an electrolyte gate voltage of \SI{1}{V}, a carrier density contrast of $\Delta n > \SI{e14}{\cm^{-2}}$ can be created at the PMMA mask boundary across only a few nanometers, as shown in the blue curve in Figure~\ref{schematic-simulation}(c), produced by finite element simulations with COMSOL-Multiphysics~\cite{COMSOL}. Plotted in Figure~\ref{schematic-simulation}(d) in the blue curve is the calculated in-plane electric field intensity across the junction, showing a maximum magnitude as high as $\si{E_{in-plane}}=\SI{600}{MV.m^{-1}}$ at the close vicinity of the mask boundary. The simulation assumes a Stern-Gouy-Chapman electrical double layer model~\cite{bard1980electrochemical} of the electrolyte ions and calculates the electric potential and the flux of ions under the influence of both ion diffusion due to the ionic concentration gradient and ion migration due to the electric field. This process is governed by the Poisson-Nernst-Planck equations. Additional details about the double layer model and parameters used in the simulation are in the Supplementary Information.

For comparison, the simulated doping contrast and the in-plane electric field are much lower in a metal-dielectric split gate, as indicated in the red curves in Figure~\ref{schematic-simulation}(c) and (d), rescaled for better visibility with factors of 5 and 10, respectively. A carrier density contrast of at most $\Delta n = 2\times\SI{e13}{\cm^{-2}}$ (1 order of magnitude lower than that with the electrolyte-PMMA-mask technique) across a length scale of $\sim\SI{100}{nm}$ is induced when a voltage of $\sim\SI{60}{V}$ is applied, corresponding to an in-plane electric field of $\si{E_{in-plane}}=\SI{13.5}{MV.m^{-1}}$ ($\sim 40$ times lower than that with the electrolyte-PMMA-mask technique). This simulation assumes a dielectric constant of 3.9 (\ce{SiO2}) and a thickness of $\SI{60}{nm}$ for the gate dielectric, and a gap of \SI{100}{nm} between the two gate electrodes, which are typical values in the literature\cite{baugher2014optoelectronic,pospischil2014solar}. The dielectric thickness and the gap width between the gate electrodes are the limiting factors for the junction sharpness. To achieve a sharpness of only a few nanometers, both the dielectric and the gap width have to be only a few nanometers in size too (see Supplementary Information). The former would result in undesirable leaking and tunneling currents and the latter is currently challenging from a fabrication standpoint.

In summary, our self-aligned electrolyte gating technique can enable nanometer-sharp junctions, and carrier density contrast and in-plane electric field orders of magnitude higher than split metal gate structures.

To implement this self-aligned electrolyte gating technique with the screening mask, the choice of material for the mask is essential. Two requirements need to be met: (1) To ensure high spatial resolution of the self-aligned local electrolyte gates, the lithographyical resolution of the mask has to be high; (2) To ensure reliable spatial selectivity and doping level control, the mask must be impenetrable to ions in the electrolyte with no leaks.

One candidate for the mask is the e-beam resist PMMA. Commonly used as a high-resolution positive-tone e-beam resist~\cite{chen1993fabrication, beaumont1981sub}, PMMA becomes a negative-tone resist when exposed at a much higher dose ($\sim\SI{20000}{\micro C.cm^{-2}}$), where it is cross-linked and transformed into graphitic nanostructures from a polymeric resist carbonization process~\cite{duan2010sub, duan2008turning}. Cross-linked PMMA allows sub-\SI{10}{nm} e-beam lithography resolution~\cite{duan2010sub}. In separate experiments, we measured negligible current between a graphene sheet covered by cross-linked PMMA ($\sim\SI{300}{nm}$ thickness) and the electrolyte gate, verifying that the mask is essentially impermeable to ions in solid polymer electrolyte \ce{PEO-LiClO4}. These cyclic voltammetry (CV) results are in the Supplementary Information.

The spatial patterning resolution of the self-aligned gates is determined by two factors: the e-beam lithography resolution of the screening mask and the Debye length of the electrolyte. As mentioned above, the e-beam resolution is $\sim\SI{10}{nm}$ and the Debye length is $\sim\SI{1}{nm}$~\cite{efetov2014towards}, so the spatial resolution for patterning local electrolyte gates using this technique is dominated by the e-beam resolution which is $\SI{10}{nm}$. The simulation in the inset of Figure~\ref{schematic-simulation}(c) indicates a well-defined carrier density modulation profile resulting from periodic local electrolyte gates with a half-pitch of $\SI{10}{nm}$. Figure~\ref{schematic-simulation}(b) shows scanning-electron-micrographs (SEMs) of two examples of PMMA mask on graphene with nanometer feature size and different geometries, including disks and ribbons.

As proof-of-principle studies, we will now apply this technique to demonstrate two different device concepts: a graphene \emph{p-n} junction and a graphene compact thermopile.

\begin{figure*}[!htbp]
\centering
\includegraphics[width=0.75\textwidth]{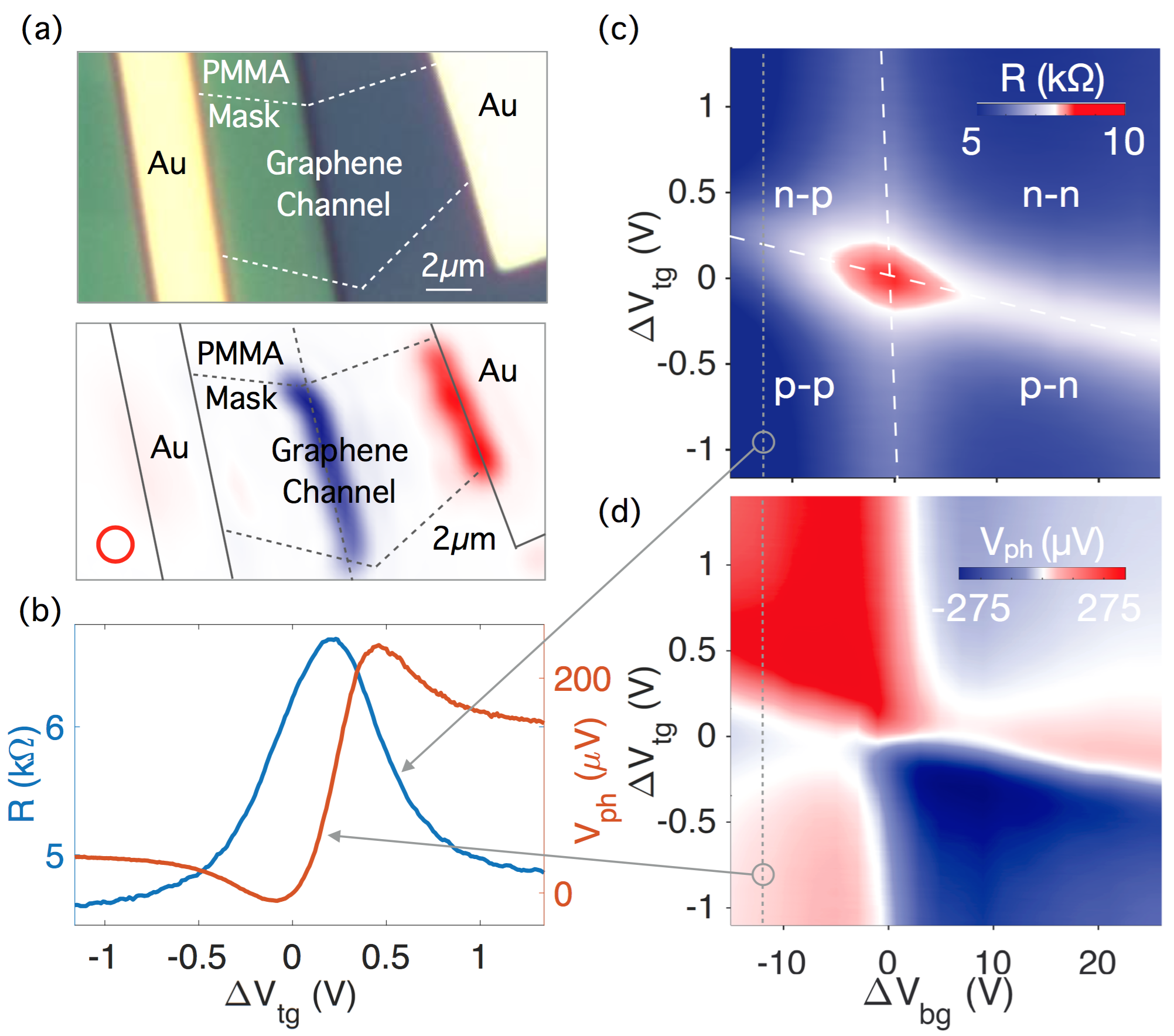}
\caption{\label{g-pn} Gate-tunable graphene \emph{p-n} junction and its photoresponse. (a) Top: Optical image of the graphene \emph{p-n} junction device. Half of the graphene channel is covered by a PMMA mask. Bottom: Photovoltage $\si{V_{ph}}$ spatial mapping of the \emph{p-n} junction device conducted on a near-infrared ($\lambda=\SI{1.55}{\micro m}$) confocal scanning microscopy setup. The red circle indicates the FWHM of the laser spot ($\sim \SI{1.7}{\micro m}$). (b) Resistance $R$ and photovoltage $\si{V_{ph}}$ as a function of $\si{V_{tg}}$, measured at $\Delta\si{V_{bg}}=\SI{-12}{V}$ (line traces for (c) and (d) along the gray dotted line). (c) Resistance $R$ as a function of $\si{V_{tg}}$ and $\si{V_{bg}}$. White dashed lines indicate charge neutrality points of the top- and bottom-gated halves of the channel, and the four regions correspond to four doping configurations of the channel. (d) Photovoltage $\si{V_{ph}}$ as a function of $\si{V_{tg}}$ and $\si{V_{bg}}$, measured at the \emph{p-n} interface.}
\end{figure*}

Figure~\ref{g-pn} shows a graphene \emph{p-n} junction and the dynamical tuning of its doping level and photoresponse. The structure of the \emph{p-n} junction device, shown in Figure~\ref{g-pn}(a) (top panel), consists of a graphene channel covered in half by a PMMA mask. The graphene is exfoliated onto a \SI{285}{nm} \ce{SiO2} substrate thermally grown on doped \ce{Si}. It is then patterned by e-beam lithography and reactive ion etching (RIE) into a channel roughly \SI{10}{\micro\meter} in length and \SI{5}{\micro\meter} in width. A pair of \ce{Cr}/\ce{Au} contacts and a PMMA mask are then defined by e-beam lithography. Solid polymer electrolyte \ce{PEO-LiClO4} is then drop-casted to cover the entire device.

Large-range doping level control of the two regions in a \emph{p-n} junction can be achieved by tuning electrolyte top gate and \ce{SiO2} back gate voltages, where $\si{V_{tg}}$ controls the mask-uncovered region and $\si{V_{bg}}$ mostly controls the mask-covered region. Figure~\ref{g-pn}(c) shows the channel resistance $R$ versus $\si{V_{tg}}$ and $\si{V_{bg}}$, showing four distinct characteristic regions that indicate gate-voltage-tunable charge density at a \emph{p-n} interface~\cite{williams2007quantum}. Two intersecting lines of high resistance (white dashed), representing charge neutrality points of the two regions respectively, divide the resistance map into four low-resistance regions: \emph{p-n}, \emph{p-p}, \emph{n-p}, and \emph{n-n}. A vertical line trace of the 2D resistance map along the dotted gray line, shown in the blue curve of Figure~\ref{g-pn}(b), exhibits distinct Dirac peak indicating modulation of graphene's Fermi level across the charge neutrality point.

Photoresponse observed at the graphene \emph{p-n} junction can also be dynamically tuned by the gate voltages. Figure~\ref{g-pn}(a) (bottom panel) shows the spatially-resolved open-circuit photovoltage map of the device under zero bias voltage across the channel, conducted on a near-infrared ($\lambda=\SI{1.55}{\micro m}$) confocal scanning microscopy setup at room temperature. As the laser excitation is scanned over the device, a large photovoltage $\si{V_{ph}}$ is observed at the self-aligned electrolyte gate defined junction. This photovoltage $\si{V_{ph}}$ at the junction as a function of $\si{V_{tg}}$ and $\si{V_{bg}}$, plotted in Figure~\ref{g-pn}(d), exhibits a distinct six-fold pattern with alternating photovoltage signs, showing a strong dependence of the photoresponse on the relative doping level of the graphene junction. This six-fold pattern indicates a photo-induced hot carrier-assisted photoresponse process at the graphene \emph{p-n} junction known as the photo-thermoelectric (PTE) effect~\cite{song2011hot}. A vertical line trace of the 2D photovoltage map along the same dotted gray line, plotted in the red curve of Figure~\ref{g-pn}(b), shows typical non-monotonic gate voltage dependence as a result of the PTE effect.

\begin{figure*}[!htbp]
\centering
\includegraphics[width=\textwidth]{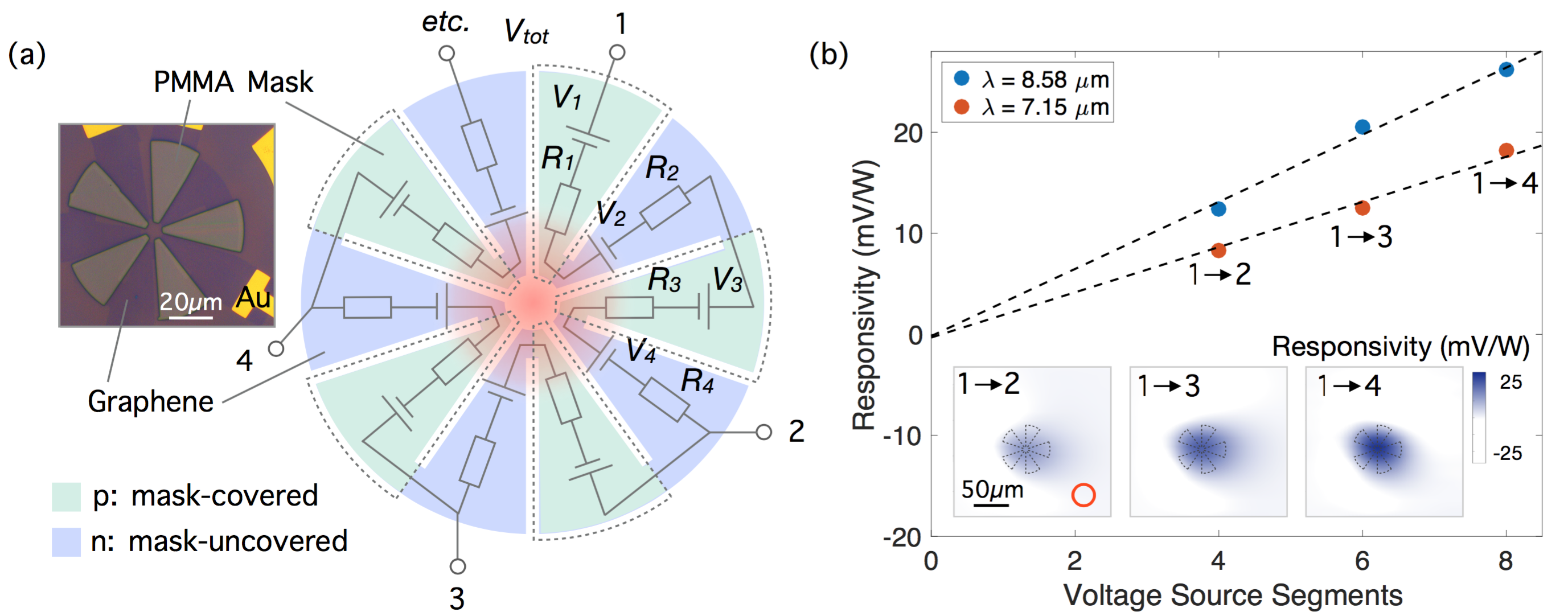}
\caption{\label{g-thermopile} Compact graphene thermopile design and its photoresponse. (a) Equivalent circuit diagram and optical image of the graphene thermopile design. Graphene is etched into a circular channel with a meandering shape, with each segment acting as a voltage source and a serial resistance. The segments are \emph{p-} or \emph{n-}doped in an alternating way to produce photovoltage in the same direction along the channel. The doping pattern is generated by covering every other graphene segments with a PMMA mask and apply appropriate $\si{V_{tg}}$ and $\si{V_{bg}}$. (b) Photovoltage responsivity of the thermopile measured at the center of the device as a function of the number of voltage source segments included in the circuit at $\lambda=\SI{8.58}{\micro m}$ and $\lambda=\SI{7.15}{\micro m}$. Data points are fitted with a linear curve indicated by the dashed line. Inset: Photovoltage responsivity spatial mappings of the thermopile measured at $\lambda=\SI{8.58}{\micro m}$, with 4, 6 and 8 voltage source segments included in the circuit, respectively. The dashed circles indicate the size of the thermopile device. The red circle indicates the FWHM of the laser spot ($\sim \SI{30}{\micro m}$).}
\end{figure*}

Next we demonstrate a compact graphene thermopile in the mid-infrared that takes advantage of the flexible gating geometries enabled by this self-aligned technique. In this design a complex doping pattern of graphene is created to enhance the photodetector's photovoltage responsivity. For PTE effect, the photovoltage generated can be expressed as $\si{V_{ph}}=\int S(n,x)\Delta T(x)\mathrm{d}x$, where $S$ is the Seebeck coefficient of graphene, a function of charge carrier density, and $\Delta T$ is the increase in electron temperature from the environment. For free-space incident light that typically has a spherical Gaussian profile, the temperature gradient points in the radial direction, so the photovoltage is maximized when it is collected radially.

The designed thermopile geometry, whose equivalent circuit diagram is illustrated in Figure~\ref{g-thermopile}(a), consists of several thermocouples connected in series whose photovoltages are all collected in the radial direction. Each graphene segment is considered a voltage source with a resistance. For the photovoltage to sum up along the meandering graphene channel, each segment is $p-$ or $n-$doped in an alternating fashion so that neighboring photovoltages point in opposite directions (Seebeck coefficient has opposite signs). The alternating doping is achieved using our gating technique by covering every other segments with the PMMA mask and applying positive $\si{V_{tg}}$ and negative $\si{V_{bg}}$ respectively. Compared to existing graphene thermopiles such as that in ref.~\cite{hsu2015graphene}, this approach eliminates the need for embedded gates and external wiring of thermocouples, enabling a compact thermopile based solely on graphene. The achievable nanoscale dimensions and the more complex geometries could lead to more efficient photovoltage collection.

Spatially-resolved open-circuit photovoltage mapping of this thermopile is conducted on a confocal scanning microscopy setup with a mid-infrared laser source at two wavelengths $\lambda=\SI{8.58}{\micro m}$ and $\lambda=\SI{7.15}{\micro m}$. The photovoltage between several sets of terminals (indicated with numbers in Figure~\ref{g-thermopile}(a)) are measured to study the individual contributions of thermocouples. As the laser spot is scanned over the device, a maximal photovoltage $\si{V_{ph}}$ is observed at the center of the thermopile, as shown in the photovoltage spatial maps in the inset of Figure~\ref{g-thermopile}(b). The spatial maximum for responsivity is then plotted as a function of the number of voltage source segments between the terminals. The well-fitted linear relation at both wavelengths confirms the summation of photovoltages from each individual segment. The maximal responsivity of this device is $\SI{26.2}{mV.W^{-1}}$. Compared to a previous graphene thermocouple with only one single \emph{p-n} junction, studied in similar conditions~\cite{herring2014photoresponse}, the carefully designed doping pattern of graphene results in an about 7 times enhancement in photovoltage responsivity. Further optimization of parameters including device dimension and the number of voltage segments can be done to achieve higher responsivity.

The flexibility to tailor the dimension and geometry of electrolyte gates on 2D materials at the nanoscale with strong doping ability expands the possibility of 2D-material-based tunable optoelectronic devices. Also, the non-destructive fabrication procedure maintains the high quality of the 2D material samples since there is no need for nanopatterning of graphene itself in harsh environments, and the cross-linked PMMA has previously been shown to have insignificant effect on the mobility of graphene~\cite{henriksen2012quantum}. Moreover, the broadband optical transparency of the PMMA mask also ensures non-interfered optical spectroscopy on the fabricated device. These are important practical considerations that can be crucial for the experimental implementation of many novel device concepts such as the compact thermopile we have demonstrated above. Plasmonics in graphene would be another example. To achieve a plasmonic resonant wavelength of $\SI{5}{\micro\meter}$ or less, graphene nanostructures need to have a feature size as small as $\SI{10}{nm}$. This typically requires direct patterning of the graphene sheet, which on this small scale would create significant edge scattering and reduce carrier mobility, limiting the quality factor of the graphene plasmonic resonances~\cite{low2014graphene}. The nanoscale electrolytic gating scheme proposed here would be an alternative and promising way of generating tunable graphene plasmons with an improved quality factor and a broader wavelength range.

To conclude, we have proposed and numerically simulated a self-aligned local electrolyte gating method of 2D materials that allows for a carrier density contrast of more than $\Delta n = \SI{e14}{\cm^{-2}}$ across a length of \SI{10}{nm} and an in-plane electric field of \SI{600}{MV.m^{-1}}. We have also developed an experimental implementation of this technique and demonstrated two different device concepts based on graphene, including a single \emph{p-n} junction with tunable doping level and photoresponse and a novel compact thermopile with enhanced photovoltage responsivity in the mid-infrared. This novel nanoscale electrolytic gating scheme is a promising and versatile experimental approach to numerous 2D material-based device concepts in tunable nanophotonics and optoelectronics, and can potentially be used for other low-dimensional material classes too.

\section*{Acknowledgement}

The research leading to these results has received funding from the U.S. Office of Naval Research (Award N00014-14-1-0349), the European Commission H2020 Programme (no.~604391, ``Graphene Flagship''), the European Research Council starting grant (307806, CarbonLight) and project GRASP (FP7-ICT-2013-613024-GRASP). C.P. was supported in part by the Stata Family Presidential Fellowship of MIT and in part by the U.S. Office of Naval Research (Award N00014-14-1-0349). D.K.E. was supported in part by an Advanced Concept Committee (ACC) program from MIT Lincoln Laboratory. S.N. was supported by the European Commission (FP7-ICT-2013-613024-GRASP) and thanks M. Batzer and R. Parret for their help in exploratory work on this subject. R.-J.S. was supported in part by the Center for Excitonics, an Energy Frontier Research Center funded by the U.S. Department of Energy, Office of Science, Office of Basic Energy Sciences under award no.~DE-SC0001088. G.G. was supported by the Swiss National Science Foundation (SNSF). Device fabrication was performed at the NanoStructures Laboratory at MIT and the Center for Nanoscale Systems, a member of the National Nanotechnology Infrastructure Network supported by the National Science Foundation (NSF). 

\bibliography{citation}

\begin{thebibliography}{44}%
\makeatletter
\providecommand \@ifxundefined [1]{%
 \@ifx{#1\undefined}
}%
\providecommand \@ifnum [1]{%
 \ifnum #1\expandafter \@firstoftwo
 \else \expandafter \@secondoftwo
 \fi
}%
\providecommand \@ifx [1]{%
 \ifx #1\expandafter \@firstoftwo
 \else \expandafter \@secondoftwo
 \fi
}%
\providecommand \natexlab [1]{#1}%
\providecommand \enquote  [1]{``#1''}%
\providecommand \bibnamefont  [1]{#1}%
\providecommand \bibfnamefont [1]{#1}%
\providecommand \citenamefont [1]{#1}%
\providecommand \href@noop [0]{\@secondoftwo}%
\providecommand \href [0]{\begingroup \@sanitize@url \@href}%
\providecommand \@href[1]{\@@startlink{#1}\@@href}%
\providecommand \@@href[1]{\endgroup#1\@@endlink}%
\providecommand \@sanitize@url [0]{\catcode `\\12\catcode `\$12\catcode
  `\&12\catcode `\#12\catcode `\^12\catcode `\_12\catcode `\%12\relax}%
\providecommand \@@startlink[1]{}%
\providecommand \@@endlink[0]{}%
\providecommand \url  [0]{\begingroup\@sanitize@url \@url }%
\providecommand \@url [1]{\endgroup\@href {#1}{\urlprefix }}%
\providecommand \urlprefix  [0]{URL }%
\providecommand \Eprint [0]{\href }%
\providecommand \doibase [0]{http://dx.doi.org/}%
\providecommand \selectlanguage [0]{\@gobble}%
\providecommand \bibinfo  [0]{\@secondoftwo}%
\providecommand \bibfield  [0]{\@secondoftwo}%
\providecommand \translation [1]{[#1]}%
\providecommand \BibitemOpen [0]{}%
\providecommand \bibitemStop [0]{}%
\providecommand \bibitemNoStop [0]{.\EOS\space}%
\providecommand \EOS [0]{\spacefactor3000\relax}%
\providecommand \BibitemShut  [1]{\csname bibitem#1\endcsname}%
\let\auto@bib@innerbib\@empty
\bibitem [{\citenamefont {Neamen}(2003)}]{neamen2003semiconductor}%
  \BibitemOpen
  \bibfield  {author} {\bibinfo {author} {\bibfnamefont {D.~A.}\ \bibnamefont
  {Neamen}},\ }\href@noop {} {\emph {\bibinfo {title} {Semiconductor physics
  and devices}}}\ (\bibinfo  {publisher} {McGraw-Hill Higher Education},\
  \bibinfo {year} {2003})\BibitemShut {NoStop}%
\bibitem [{\citenamefont {Chuang}(2012)}]{chuang2012physics}%
  \BibitemOpen
  \bibfield  {author} {\bibinfo {author} {\bibfnamefont {S.~L.}\ \bibnamefont
  {Chuang}},\ }\href@noop {} {\emph {\bibinfo {title} {Physics of photonic
  devices}}},\ Vol.~\bibinfo {volume} {80}\ (\bibinfo  {publisher} {John Wiley
  \& Sons},\ \bibinfo {year} {2012})\BibitemShut {NoStop}%
\bibitem [{\citenamefont {Ferrari}\ \emph {et~al.}(2015)\citenamefont
  {Ferrari}, \citenamefont {Bonaccorso}, \citenamefont {Falko}, \citenamefont
  {Novoselov}, \citenamefont {Roche}, \citenamefont {B{\o}ggild}, \citenamefont
  {Borini}, \citenamefont {Koppens}, \citenamefont {Palermo}, \citenamefont
  {Pugno}, \citenamefont {Garrido}, \citenamefont {Sordan}, \citenamefont
  {Bianco}, \citenamefont {Ballerini}, \citenamefont {Prato}, \citenamefont
  {Lidorikis}, \citenamefont {Kivioja}, \citenamefont {Marinelli},
  \citenamefont {Ryh{\"{a}}nen}, \citenamefont {Morpurgo}, \citenamefont
  {Coleman}, \citenamefont {Nicolosi}, \citenamefont {Colombo}, \citenamefont
  {Fert}, \citenamefont {Garcia-Hernandez}, \citenamefont {Bachtold},
  \citenamefont {Schneider}, \citenamefont {Guinea}, \citenamefont {Dekker},
  \citenamefont {Barbone}, \citenamefont {Galiotis}, \citenamefont
  {Grigorenko}, \citenamefont {Konstantatos}, \citenamefont {Kis},
  \citenamefont {Katsnelson}, \citenamefont {Beenakker}, \citenamefont
  {Vandersypen}, \citenamefont {Loiseau}, \citenamefont {Morandi},
  \citenamefont {Neumaier}, \citenamefont {Treossi}, \citenamefont
  {Pellegrini}, \citenamefont {Polini}, \citenamefont {Tredicucci},
  \citenamefont {Williams}, \citenamefont {Hong}, \citenamefont {Ahn},
  \citenamefont {Kim}, \citenamefont {Zirath}, \citenamefont {van Wees},
  \citenamefont {van~der Zant}, \citenamefont {Occhipinti}, \citenamefont {{Di
  Matteo}}, \citenamefont {Kinloch}, \citenamefont {Seyller}, \citenamefont
  {Quesnel}, \citenamefont {Feng}, \citenamefont {Teo}, \citenamefont
  {Rupesinghe}, \citenamefont {Hakonen}, \citenamefont {Neil}, \citenamefont
  {Tannock}, \citenamefont {L{\"{o}}fwander},\ and\ \citenamefont
  {Kinaret}}]{ferrari2015science}%
  \BibitemOpen
  \bibfield  {author} {\bibinfo {author} {\bibfnamefont {A.~C.}\ \bibnamefont
  {Ferrari}}, \bibinfo {author} {\bibfnamefont {F.}~\bibnamefont {Bonaccorso}},
  \bibinfo {author} {\bibfnamefont {V.}~\bibnamefont {Falko}}, \bibinfo
  {author} {\bibfnamefont {K.~S.}\ \bibnamefont {Novoselov}}, \bibinfo {author}
  {\bibfnamefont {S.}~\bibnamefont {Roche}}, \bibinfo {author} {\bibfnamefont
  {P.}~\bibnamefont {B{\o}ggild}}, \bibinfo {author} {\bibfnamefont
  {S.}~\bibnamefont {Borini}}, \bibinfo {author} {\bibfnamefont
  {F.}~\bibnamefont {Koppens}}, \bibinfo {author} {\bibfnamefont
  {V.}~\bibnamefont {Palermo}}, \bibinfo {author} {\bibfnamefont
  {N.}~\bibnamefont {Pugno}}, \bibinfo {author} {\bibfnamefont {J.~a.}\
  \bibnamefont {Garrido}}, \bibinfo {author} {\bibfnamefont {R.}~\bibnamefont
  {Sordan}}, \bibinfo {author} {\bibfnamefont {A.}~\bibnamefont {Bianco}},
  \bibinfo {author} {\bibfnamefont {L.}~\bibnamefont {Ballerini}}, \bibinfo
  {author} {\bibfnamefont {M.}~\bibnamefont {Prato}}, \bibinfo {author}
  {\bibfnamefont {E.}~\bibnamefont {Lidorikis}}, \bibinfo {author}
  {\bibfnamefont {J.}~\bibnamefont {Kivioja}}, \bibinfo {author} {\bibfnamefont
  {C.}~\bibnamefont {Marinelli}}, \bibinfo {author} {\bibfnamefont
  {T.}~\bibnamefont {Ryh{\"{a}}nen}}, \bibinfo {author} {\bibfnamefont
  {A.}~\bibnamefont {Morpurgo}}, \bibinfo {author} {\bibfnamefont {J.~N.}\
  \bibnamefont {Coleman}}, \bibinfo {author} {\bibfnamefont {V.}~\bibnamefont
  {Nicolosi}}, \bibinfo {author} {\bibfnamefont {L.}~\bibnamefont {Colombo}},
  \bibinfo {author} {\bibfnamefont {A.}~\bibnamefont {Fert}}, \bibinfo {author}
  {\bibfnamefont {M.}~\bibnamefont {Garcia-Hernandez}}, \bibinfo {author}
  {\bibfnamefont {A.}~\bibnamefont {Bachtold}}, \bibinfo {author}
  {\bibfnamefont {G.~F.}\ \bibnamefont {Schneider}}, \bibinfo {author}
  {\bibfnamefont {F.}~\bibnamefont {Guinea}}, \bibinfo {author} {\bibfnamefont
  {C.}~\bibnamefont {Dekker}}, \bibinfo {author} {\bibfnamefont
  {M.}~\bibnamefont {Barbone}}, \bibinfo {author} {\bibfnamefont
  {C.}~\bibnamefont {Galiotis}}, \bibinfo {author} {\bibfnamefont
  {A.}~\bibnamefont {Grigorenko}}, \bibinfo {author} {\bibfnamefont
  {G.}~\bibnamefont {Konstantatos}}, \bibinfo {author} {\bibfnamefont
  {A.}~\bibnamefont {Kis}}, \bibinfo {author} {\bibfnamefont {M.}~\bibnamefont
  {Katsnelson}}, \bibinfo {author} {\bibfnamefont {C.~W.~J.}\ \bibnamefont
  {Beenakker}}, \bibinfo {author} {\bibfnamefont {L.}~\bibnamefont
  {Vandersypen}}, \bibinfo {author} {\bibfnamefont {A.}~\bibnamefont
  {Loiseau}}, \bibinfo {author} {\bibfnamefont {V.}~\bibnamefont {Morandi}},
  \bibinfo {author} {\bibfnamefont {D.}~\bibnamefont {Neumaier}}, \bibinfo
  {author} {\bibfnamefont {E.}~\bibnamefont {Treossi}}, \bibinfo {author}
  {\bibfnamefont {V.}~\bibnamefont {Pellegrini}}, \bibinfo {author}
  {\bibfnamefont {M.}~\bibnamefont {Polini}}, \bibinfo {author} {\bibfnamefont
  {A.}~\bibnamefont {Tredicucci}}, \bibinfo {author} {\bibfnamefont {G.~M.}\
  \bibnamefont {Williams}}, \bibinfo {author} {\bibfnamefont {B.~H.}\
  \bibnamefont {Hong}}, \bibinfo {author} {\bibfnamefont {J.~H.}\ \bibnamefont
  {Ahn}}, \bibinfo {author} {\bibfnamefont {J.~M.}\ \bibnamefont {Kim}},
  \bibinfo {author} {\bibfnamefont {H.}~\bibnamefont {Zirath}}, \bibinfo
  {author} {\bibfnamefont {B.~J.}\ \bibnamefont {van Wees}}, \bibinfo {author}
  {\bibfnamefont {H.}~\bibnamefont {van~der Zant}}, \bibinfo {author}
  {\bibfnamefont {L.}~\bibnamefont {Occhipinti}}, \bibinfo {author}
  {\bibfnamefont {A.}~\bibnamefont {{Di Matteo}}}, \bibinfo {author}
  {\bibfnamefont {I.~a.}\ \bibnamefont {Kinloch}}, \bibinfo {author}
  {\bibfnamefont {T.}~\bibnamefont {Seyller}}, \bibinfo {author} {\bibfnamefont
  {E.}~\bibnamefont {Quesnel}}, \bibinfo {author} {\bibfnamefont
  {X.}~\bibnamefont {Feng}}, \bibinfo {author} {\bibfnamefont {K.}~\bibnamefont
  {Teo}}, \bibinfo {author} {\bibfnamefont {N.}~\bibnamefont {Rupesinghe}},
  \bibinfo {author} {\bibfnamefont {P.}~\bibnamefont {Hakonen}}, \bibinfo
  {author} {\bibfnamefont {S.~R.~T.}\ \bibnamefont {Neil}}, \bibinfo {author}
  {\bibfnamefont {Q.}~\bibnamefont {Tannock}}, \bibinfo {author} {\bibfnamefont
  {T.}~\bibnamefont {L{\"{o}}fwander}}, \ and\ \bibinfo {author} {\bibfnamefont
  {J.}~\bibnamefont {Kinaret}},\ }\bibfield  {title} {\enquote {\bibinfo
  {title} {Science and technology roadmap for graphene, related two-dimensional
  crystals, and hybrid systems},}\ }\href@noop {} {\bibfield  {journal}
  {\bibinfo  {journal} {Nanoscale}\ }\textbf {\bibinfo {volume} {7}},\ \bibinfo
  {pages} {4598--4810} (\bibinfo {year} {2015})}\BibitemShut {NoStop}%
\bibitem [{\citenamefont {Bao}\ and\ \citenamefont
  {Loh}(2012)}]{bao2012graphene}%
  \BibitemOpen
  \bibfield  {author} {\bibinfo {author} {\bibfnamefont {Q.}~\bibnamefont
  {Bao}}\ and\ \bibinfo {author} {\bibfnamefont {K.~P.}\ \bibnamefont {Loh}},\
  }\bibfield  {title} {\enquote {\bibinfo {title} {Graphene photonics,
  plasmonics, and broadband optoelectronic devices},}\ }\href@noop {}
  {\bibfield  {journal} {\bibinfo  {journal} {ACS nano}\ }\textbf {\bibinfo
  {volume} {6}},\ \bibinfo {pages} {3677--3694} (\bibinfo {year}
  {2012})}\BibitemShut {NoStop}%
\bibitem [{\citenamefont {Avouris}\ and\ \citenamefont
  {Freitag}(2014)}]{avouris2014graphene}%
  \BibitemOpen
  \bibfield  {author} {\bibinfo {author} {\bibfnamefont {P.}~\bibnamefont
  {Avouris}}\ and\ \bibinfo {author} {\bibfnamefont {M.}~\bibnamefont
  {Freitag}},\ }\bibfield  {title} {\enquote {\bibinfo {title} {Graphene
  photonics, plasmonics, and optoelectronics},}\ }\href@noop {} {\bibfield
  {journal} {\bibinfo  {journal} {IEEE Journal of selected topics in quantum
  electronics}\ }\textbf {\bibinfo {volume} {1}},\ \bibinfo {pages} {6000112}
  (\bibinfo {year} {2014})}\BibitemShut {NoStop}%
\bibitem [{\citenamefont {Koppens}\ \emph {et~al.}(2014)\citenamefont
  {Koppens}, \citenamefont {Mueller}, \citenamefont {Avouris}, \citenamefont
  {Ferrari}, \citenamefont {Vitiello},\ and\ \citenamefont
  {Polini}}]{koppens2014photodetectors}%
  \BibitemOpen
  \bibfield  {author} {\bibinfo {author} {\bibfnamefont {F.}~\bibnamefont
  {Koppens}}, \bibinfo {author} {\bibfnamefont {T.}~\bibnamefont {Mueller}},
  \bibinfo {author} {\bibfnamefont {P.}~\bibnamefont {Avouris}}, \bibinfo
  {author} {\bibfnamefont {A.}~\bibnamefont {Ferrari}}, \bibinfo {author}
  {\bibfnamefont {M.}~\bibnamefont {Vitiello}}, \ and\ \bibinfo {author}
  {\bibfnamefont {M.}~\bibnamefont {Polini}},\ }\bibfield  {title} {\enquote
  {\bibinfo {title} {Photodetectors based on graphene, other two-dimensional
  materials and hybrid systems},}\ }\href@noop {} {\bibfield  {journal}
  {\bibinfo  {journal} {Nature nanotechnology}\ }\textbf {\bibinfo {volume}
  {9}},\ \bibinfo {pages} {780--793} (\bibinfo {year} {2014})}\BibitemShut
  {NoStop}%
\bibitem [{\citenamefont {Wang}\ \emph {et~al.}(2012)\citenamefont {Wang},
  \citenamefont {Kalantar-Zadeh}, \citenamefont {Kis}, \citenamefont
  {Coleman},\ and\ \citenamefont {Strano}}]{wang2012electronics}%
  \BibitemOpen
  \bibfield  {author} {\bibinfo {author} {\bibfnamefont {Q.~H.}\ \bibnamefont
  {Wang}}, \bibinfo {author} {\bibfnamefont {K.}~\bibnamefont
  {Kalantar-Zadeh}}, \bibinfo {author} {\bibfnamefont {A.}~\bibnamefont {Kis}},
  \bibinfo {author} {\bibfnamefont {J.~N.}\ \bibnamefont {Coleman}}, \ and\
  \bibinfo {author} {\bibfnamefont {M.~S.}\ \bibnamefont {Strano}},\ }\bibfield
   {title} {\enquote {\bibinfo {title} {Electronics and optoelectronics of
  two-dimensional transition metal dichalcogenides},}\ }\href@noop {}
  {\bibfield  {journal} {\bibinfo  {journal} {Nature nanotechnology}\ }\textbf
  {\bibinfo {volume} {7}},\ \bibinfo {pages} {699--712} (\bibinfo {year}
  {2012})}\BibitemShut {NoStop}%
\bibitem [{\citenamefont {Britnell}\ \emph {et~al.}(2013)\citenamefont
  {Britnell}, \citenamefont {Ribeiro}, \citenamefont {Eckmann}, \citenamefont
  {Jalil}, \citenamefont {Belle}, \citenamefont {Mishchenko}, \citenamefont
  {Kim}, \citenamefont {Gorbachev}, \citenamefont {Georgiou}, \citenamefont
  {Morozov}, \citenamefont {Grigorenko}, \citenamefont {Geim}, \citenamefont
  {Casiraghi}, \citenamefont {Neto},\ and\ \citenamefont
  {Novoselov}}]{britnell2013strong}%
  \BibitemOpen
  \bibfield  {author} {\bibinfo {author} {\bibfnamefont {L.}~\bibnamefont
  {Britnell}}, \bibinfo {author} {\bibfnamefont {R.~M.}\ \bibnamefont
  {Ribeiro}}, \bibinfo {author} {\bibfnamefont {A.}~\bibnamefont {Eckmann}},
  \bibinfo {author} {\bibfnamefont {R.}~\bibnamefont {Jalil}}, \bibinfo
  {author} {\bibfnamefont {B.~D.}\ \bibnamefont {Belle}}, \bibinfo {author}
  {\bibfnamefont {A.}~\bibnamefont {Mishchenko}}, \bibinfo {author}
  {\bibfnamefont {Y.-J.}\ \bibnamefont {Kim}}, \bibinfo {author} {\bibfnamefont
  {R.~V.}\ \bibnamefont {Gorbachev}}, \bibinfo {author} {\bibfnamefont
  {T.}~\bibnamefont {Georgiou}}, \bibinfo {author} {\bibfnamefont {S.~V.}\
  \bibnamefont {Morozov}}, \bibinfo {author} {\bibfnamefont {A.~N.}\
  \bibnamefont {Grigorenko}}, \bibinfo {author} {\bibfnamefont {A.~K.}\
  \bibnamefont {Geim}}, \bibinfo {author} {\bibfnamefont {C.}~\bibnamefont
  {Casiraghi}}, \bibinfo {author} {\bibfnamefont {A.~H.~C.}\ \bibnamefont
  {Neto}}, \ and\ \bibinfo {author} {\bibfnamefont {K.~S.}\ \bibnamefont
  {Novoselov}},\ }\bibfield  {title} {\enquote {\bibinfo {title} {Strong
  light-matter interactions in heterostructures of atomically thin films},}\
  }\href@noop {} {\bibfield  {journal} {\bibinfo  {journal} {Science}\ }\textbf
  {\bibinfo {volume} {340}},\ \bibinfo {pages} {1311--1314} (\bibinfo {year}
  {2013})}\BibitemShut {NoStop}%
\bibitem [{\citenamefont {Liu}\ \emph {et~al.}(2014)\citenamefont {Liu},
  \citenamefont {Chang}, \citenamefont {Norris},\ and\ \citenamefont
  {Zhong}}]{liu2014graphene}%
  \BibitemOpen
  \bibfield  {author} {\bibinfo {author} {\bibfnamefont {C.-H.}\ \bibnamefont
  {Liu}}, \bibinfo {author} {\bibfnamefont {Y.-C.}\ \bibnamefont {Chang}},
  \bibinfo {author} {\bibfnamefont {T.~B.}\ \bibnamefont {Norris}}, \ and\
  \bibinfo {author} {\bibfnamefont {Z.}~\bibnamefont {Zhong}},\ }\bibfield
  {title} {\enquote {\bibinfo {title} {Graphene photodetectors with
  ultra-broadband and high responsivity at room temperature},}\ }\href@noop {}
  {\bibfield  {journal} {\bibinfo  {journal} {Nature nanotechnology}\ }\textbf
  {\bibinfo {volume} {9}},\ \bibinfo {pages} {273--278} (\bibinfo {year}
  {2014})}\BibitemShut {NoStop}%
\bibitem [{\citenamefont {Sassi}\ \emph {et~al.}(2016)\citenamefont {Sassi},
  \citenamefont {Parret}, \citenamefont {Nanot}, \citenamefont {Bruna},
  \citenamefont {Borini}, \citenamefont {Milana}, \citenamefont {De~Fazio},
  \citenamefont {Zhuang}, \citenamefont {Lidorikis}, \citenamefont {Koppens}
  \emph {et~al.}}]{sassi2016graphene}%
  \BibitemOpen
  \bibfield  {author} {\bibinfo {author} {\bibfnamefont {U.}~\bibnamefont
  {Sassi}}, \bibinfo {author} {\bibfnamefont {R.}~\bibnamefont {Parret}},
  \bibinfo {author} {\bibfnamefont {S.}~\bibnamefont {Nanot}}, \bibinfo
  {author} {\bibfnamefont {M.}~\bibnamefont {Bruna}}, \bibinfo {author}
  {\bibfnamefont {S.}~\bibnamefont {Borini}}, \bibinfo {author} {\bibfnamefont
  {S.}~\bibnamefont {Milana}}, \bibinfo {author} {\bibfnamefont
  {D.}~\bibnamefont {De~Fazio}}, \bibinfo {author} {\bibfnamefont
  {Z.}~\bibnamefont {Zhuang}}, \bibinfo {author} {\bibfnamefont
  {E.}~\bibnamefont {Lidorikis}}, \bibinfo {author} {\bibfnamefont
  {F.}~\bibnamefont {Koppens}},  \emph {et~al.},\ }\bibfield  {title} {\enquote
  {\bibinfo {title} {Graphene-based, mid-infrared, room-temperature
  pyroelectric bolometers with ultrahigh temperature coefficient of
  resistance},}\ }\href@noop {} {\bibfield  {journal} {\bibinfo  {journal}
  {arXiv preprint arXiv:1608.00569}\ } (\bibinfo {year} {2016})}\BibitemShut
  {NoStop}%
\bibitem [{\citenamefont {Roy}\ \emph {et~al.}(2015)\citenamefont {Roy},
  \citenamefont {Tosun}, \citenamefont {Cao}, \citenamefont {Fang},
  \citenamefont {Lien}, \citenamefont {Zhao}, \citenamefont {Chen},
  \citenamefont {Chueh}, \citenamefont {Guo},\ and\ \citenamefont
  {Javey}}]{roy2015dual}%
  \BibitemOpen
  \bibfield  {author} {\bibinfo {author} {\bibfnamefont {T.}~\bibnamefont
  {Roy}}, \bibinfo {author} {\bibfnamefont {M.}~\bibnamefont {Tosun}}, \bibinfo
  {author} {\bibfnamefont {X.}~\bibnamefont {Cao}}, \bibinfo {author}
  {\bibfnamefont {H.}~\bibnamefont {Fang}}, \bibinfo {author} {\bibfnamefont
  {D.-H.}\ \bibnamefont {Lien}}, \bibinfo {author} {\bibfnamefont
  {P.}~\bibnamefont {Zhao}}, \bibinfo {author} {\bibfnamefont {Y.-Z.}\
  \bibnamefont {Chen}}, \bibinfo {author} {\bibfnamefont {Y.-L.}\ \bibnamefont
  {Chueh}}, \bibinfo {author} {\bibfnamefont {J.}~\bibnamefont {Guo}}, \ and\
  \bibinfo {author} {\bibfnamefont {A.}~\bibnamefont {Javey}},\ }\bibfield
  {title} {\enquote {\bibinfo {title} {Dual-gated mos2/wse2 van der waals
  tunnel diodes and transistors},}\ }\href@noop {} {\bibfield  {journal}
  {\bibinfo  {journal} {Acs Nano}\ }\textbf {\bibinfo {volume} {9}},\ \bibinfo
  {pages} {2071--2079} (\bibinfo {year} {2015})}\BibitemShut {NoStop}%
\bibitem [{\citenamefont {Cheianov}, \citenamefont {Fal'ko},\ and\
  \citenamefont {Altshuler}(2007)}]{cheianov2007focusing}%
  \BibitemOpen
  \bibfield  {author} {\bibinfo {author} {\bibfnamefont {V.~V.}\ \bibnamefont
  {Cheianov}}, \bibinfo {author} {\bibfnamefont {V.}~\bibnamefont {Fal'ko}}, \
  and\ \bibinfo {author} {\bibfnamefont {B.}~\bibnamefont {Altshuler}},\
  }\bibfield  {title} {\enquote {\bibinfo {title} {The focusing of electron
  flow and a veselago lens in graphene pn junctions},}\ }\href@noop {}
  {\bibfield  {journal} {\bibinfo  {journal} {Science}\ }\textbf {\bibinfo
  {volume} {315}},\ \bibinfo {pages} {1252--1255} (\bibinfo {year}
  {2007})}\BibitemShut {NoStop}%
\bibitem [{\citenamefont {Shang}\ \emph {et~al.}(2011)\citenamefont {Shang},
  \citenamefont {He}, \citenamefont {Wang}, \citenamefont {Li}, \citenamefont
  {Zhu}, \citenamefont {Niu},\ and\ \citenamefont {Fu}}]{shang2011effect}%
  \BibitemOpen
  \bibfield  {author} {\bibinfo {author} {\bibfnamefont {X.}~\bibnamefont
  {Shang}}, \bibinfo {author} {\bibfnamefont {J.}~\bibnamefont {He}}, \bibinfo
  {author} {\bibfnamefont {H.}~\bibnamefont {Wang}}, \bibinfo {author}
  {\bibfnamefont {M.}~\bibnamefont {Li}}, \bibinfo {author} {\bibfnamefont
  {Y.}~\bibnamefont {Zhu}}, \bibinfo {author} {\bibfnamefont {Z.}~\bibnamefont
  {Niu}}, \ and\ \bibinfo {author} {\bibfnamefont {Y.}~\bibnamefont {Fu}},\
  }\bibfield  {title} {\enquote {\bibinfo {title} {Effect of built-in electric
  field in photovoltaic inas quantum dot embedded gaas solar cell},}\
  }\href@noop {} {\bibfield  {journal} {\bibinfo  {journal} {Applied Physics
  A}\ }\textbf {\bibinfo {volume} {103}},\ \bibinfo {pages} {335--341}
  (\bibinfo {year} {2011})}\BibitemShut {NoStop}%
\bibitem [{\citenamefont {Konstantatos}\ and\ \citenamefont
  {Sargent}(2010)}]{konstantatos2010nanostructured}%
  \BibitemOpen
  \bibfield  {author} {\bibinfo {author} {\bibfnamefont {G.}~\bibnamefont
  {Konstantatos}}\ and\ \bibinfo {author} {\bibfnamefont {E.~H.}\ \bibnamefont
  {Sargent}},\ }\bibfield  {title} {\enquote {\bibinfo {title} {Nanostructured
  materials for photon detection},}\ }\href@noop {} {\bibfield  {journal}
  {\bibinfo  {journal} {Nature nanotechnology}\ }\textbf {\bibinfo {volume}
  {5}},\ \bibinfo {pages} {391--400} (\bibinfo {year} {2010})}\BibitemShut
  {NoStop}%
\bibitem [{\citenamefont {Thongrattanasiri}, \citenamefont {Koppens},\ and\
  \citenamefont {de~Abajo}(2012)}]{thongrattanasiri2012complete}%
  \BibitemOpen
  \bibfield  {author} {\bibinfo {author} {\bibfnamefont {S.}~\bibnamefont
  {Thongrattanasiri}}, \bibinfo {author} {\bibfnamefont {F.~H.}\ \bibnamefont
  {Koppens}}, \ and\ \bibinfo {author} {\bibfnamefont {F.~J.~G.}\ \bibnamefont
  {de~Abajo}},\ }\bibfield  {title} {\enquote {\bibinfo {title} {Complete
  optical absorption in periodically patterned graphene},}\ }\href@noop {}
  {\bibfield  {journal} {\bibinfo  {journal} {Physical review letters}\
  }\textbf {\bibinfo {volume} {108}},\ \bibinfo {pages} {047401} (\bibinfo
  {year} {2012})}\BibitemShut {NoStop}%
\bibitem [{\citenamefont {Ye}\ \emph {et~al.}(2015)\citenamefont {Ye},
  \citenamefont {Zhu}, \citenamefont {Xu}, \citenamefont {Yuan},\ and\
  \citenamefont {Qin}}]{ye2015electrically}%
  \BibitemOpen
  \bibfield  {author} {\bibinfo {author} {\bibfnamefont {C.}~\bibnamefont
  {Ye}}, \bibinfo {author} {\bibfnamefont {Z.}~\bibnamefont {Zhu}}, \bibinfo
  {author} {\bibfnamefont {W.}~\bibnamefont {Xu}}, \bibinfo {author}
  {\bibfnamefont {X.}~\bibnamefont {Yuan}}, \ and\ \bibinfo {author}
  {\bibfnamefont {S.}~\bibnamefont {Qin}},\ }\bibfield  {title} {\enquote
  {\bibinfo {title} {Electrically tunable absorber based on nonstructured
  graphene},}\ }\href@noop {} {\bibfield  {journal} {\bibinfo  {journal}
  {Journal of Optics}\ }\textbf {\bibinfo {volume} {17}},\ \bibinfo {pages}
  {125009} (\bibinfo {year} {2015})}\BibitemShut {NoStop}%
\bibitem [{\citenamefont {Vakil}\ and\ \citenamefont
  {Engheta}(2011)}]{vakil2011transformation}%
  \BibitemOpen
  \bibfield  {author} {\bibinfo {author} {\bibfnamefont {A.}~\bibnamefont
  {Vakil}}\ and\ \bibinfo {author} {\bibfnamefont {N.}~\bibnamefont
  {Engheta}},\ }\bibfield  {title} {\enquote {\bibinfo {title} {Transformation
  optics using graphene},}\ }\href@noop {} {\bibfield  {journal} {\bibinfo
  {journal} {Science}\ }\textbf {\bibinfo {volume} {332}},\ \bibinfo {pages}
  {1291--1294} (\bibinfo {year} {2011})}\BibitemShut {NoStop}%
\bibitem [{\citenamefont {Yankowitz}\ \emph {et~al.}(2012)\citenamefont
  {Yankowitz}, \citenamefont {Xue}, \citenamefont {Cormode}, \citenamefont
  {Sanchez-Yamagishi}, \citenamefont {Watanabe}, \citenamefont {Taniguchi},
  \citenamefont {Jarillo-Herrero}, \citenamefont {Jacquod},\ and\ \citenamefont
  {LeRoy}}]{yankowitz2012emergence}%
  \BibitemOpen
  \bibfield  {author} {\bibinfo {author} {\bibfnamefont {M.}~\bibnamefont
  {Yankowitz}}, \bibinfo {author} {\bibfnamefont {J.}~\bibnamefont {Xue}},
  \bibinfo {author} {\bibfnamefont {D.}~\bibnamefont {Cormode}}, \bibinfo
  {author} {\bibfnamefont {J.~D.}\ \bibnamefont {Sanchez-Yamagishi}}, \bibinfo
  {author} {\bibfnamefont {K.}~\bibnamefont {Watanabe}}, \bibinfo {author}
  {\bibfnamefont {T.}~\bibnamefont {Taniguchi}}, \bibinfo {author}
  {\bibfnamefont {P.}~\bibnamefont {Jarillo-Herrero}}, \bibinfo {author}
  {\bibfnamefont {P.}~\bibnamefont {Jacquod}}, \ and\ \bibinfo {author}
  {\bibfnamefont {B.~J.}\ \bibnamefont {LeRoy}},\ }\bibfield  {title} {\enquote
  {\bibinfo {title} {Emergence of superlattice dirac points in graphene on
  hexagonal boron nitride},}\ }\href@noop {} {\bibfield  {journal} {\bibinfo
  {journal} {Nature Physics}\ }\textbf {\bibinfo {volume} {8}},\ \bibinfo
  {pages} {382--386} (\bibinfo {year} {2012})}\BibitemShut {NoStop}%
\bibitem [{\citenamefont {Kang}\ \emph {et~al.}(2013)\citenamefont {Kang},
  \citenamefont {Li}, \citenamefont {Li}, \citenamefont {Xia},\ and\
  \citenamefont {Wang}}]{kang2013electronic}%
  \BibitemOpen
  \bibfield  {author} {\bibinfo {author} {\bibfnamefont {J.}~\bibnamefont
  {Kang}}, \bibinfo {author} {\bibfnamefont {J.}~\bibnamefont {Li}}, \bibinfo
  {author} {\bibfnamefont {S.-S.}\ \bibnamefont {Li}}, \bibinfo {author}
  {\bibfnamefont {J.-B.}\ \bibnamefont {Xia}}, \ and\ \bibinfo {author}
  {\bibfnamefont {L.-W.}\ \bibnamefont {Wang}},\ }\bibfield  {title} {\enquote
  {\bibinfo {title} {Electronic structural moire pattern effects on mos2/mose2
  2d heterostructures},}\ }\href@noop {} {\bibfield  {journal} {\bibinfo
  {journal} {Nano letters}\ }\textbf {\bibinfo {volume} {13}},\ \bibinfo
  {pages} {5485--5490} (\bibinfo {year} {2013})}\BibitemShut {NoStop}%
\bibitem [{\citenamefont {Park}\ \emph
  {et~al.}(2008{\natexlab{a}})\citenamefont {Park}, \citenamefont {Yang},
  \citenamefont {Son}, \citenamefont {Cohen},\ and\ \citenamefont
  {Louie}}]{park2008anisotropic}%
  \BibitemOpen
  \bibfield  {author} {\bibinfo {author} {\bibfnamefont {C.-H.}\ \bibnamefont
  {Park}}, \bibinfo {author} {\bibfnamefont {L.}~\bibnamefont {Yang}}, \bibinfo
  {author} {\bibfnamefont {Y.-W.}\ \bibnamefont {Son}}, \bibinfo {author}
  {\bibfnamefont {M.~L.}\ \bibnamefont {Cohen}}, \ and\ \bibinfo {author}
  {\bibfnamefont {S.~G.}\ \bibnamefont {Louie}},\ }\bibfield  {title} {\enquote
  {\bibinfo {title} {Anisotropic behaviours of massless dirac fermions in
  graphene under periodic potentials},}\ }\href@noop {} {\bibfield  {journal}
  {\bibinfo  {journal} {Nature Physics}\ }\textbf {\bibinfo {volume} {4}},\
  \bibinfo {pages} {213--217} (\bibinfo {year}
  {2008}{\natexlab{a}})}\BibitemShut {NoStop}%
\bibitem [{\citenamefont {Park}\ \emph
  {et~al.}(2008{\natexlab{b}})\citenamefont {Park}, \citenamefont {Son},
  \citenamefont {Yang}, \citenamefont {Cohen},\ and\ \citenamefont
  {Louie}}]{park2008electron}%
  \BibitemOpen
  \bibfield  {author} {\bibinfo {author} {\bibfnamefont {C.-H.}\ \bibnamefont
  {Park}}, \bibinfo {author} {\bibfnamefont {Y.-W.}\ \bibnamefont {Son}},
  \bibinfo {author} {\bibfnamefont {L.}~\bibnamefont {Yang}}, \bibinfo {author}
  {\bibfnamefont {M.~L.}\ \bibnamefont {Cohen}}, \ and\ \bibinfo {author}
  {\bibfnamefont {S.~G.}\ \bibnamefont {Louie}},\ }\bibfield  {title} {\enquote
  {\bibinfo {title} {Electron beam supercollimation in graphene
  superlattices},}\ }\href@noop {} {\bibfield  {journal} {\bibinfo  {journal}
  {Nano letters}\ }\textbf {\bibinfo {volume} {8}},\ \bibinfo {pages}
  {2920--2924} (\bibinfo {year} {2008}{\natexlab{b}})}\BibitemShut {NoStop}%
\bibitem [{\citenamefont {Baugher}\ \emph {et~al.}(2014)\citenamefont
  {Baugher}, \citenamefont {Churchill}, \citenamefont {Yang},\ and\
  \citenamefont {Jarillo-Herrero}}]{baugher2014optoelectronic}%
  \BibitemOpen
  \bibfield  {author} {\bibinfo {author} {\bibfnamefont {B.~W.}\ \bibnamefont
  {Baugher}}, \bibinfo {author} {\bibfnamefont {H.~O.}\ \bibnamefont
  {Churchill}}, \bibinfo {author} {\bibfnamefont {Y.}~\bibnamefont {Yang}}, \
  and\ \bibinfo {author} {\bibfnamefont {P.}~\bibnamefont {Jarillo-Herrero}},\
  }\bibfield  {title} {\enquote {\bibinfo {title} {Optoelectronic devices based
  on electrically tunable pn diodes in a monolayer dichalcogenide},}\
  }\href@noop {} {\bibfield  {journal} {\bibinfo  {journal} {Nature
  nanotechnology}\ }\textbf {\bibinfo {volume} {9}},\ \bibinfo {pages}
  {262--267} (\bibinfo {year} {2014})}\BibitemShut {NoStop}%
\bibitem [{\citenamefont {Pospischil}, \citenamefont {Furchi},\ and\
  \citenamefont {Mueller}(2014)}]{pospischil2014solar}%
  \BibitemOpen
  \bibfield  {author} {\bibinfo {author} {\bibfnamefont {A.}~\bibnamefont
  {Pospischil}}, \bibinfo {author} {\bibfnamefont {M.~M.}\ \bibnamefont
  {Furchi}}, \ and\ \bibinfo {author} {\bibfnamefont {T.}~\bibnamefont
  {Mueller}},\ }\bibfield  {title} {\enquote {\bibinfo {title} {Solar-energy
  conversion and light emission in an atomic monolayer pn diode},}\ }\href@noop
  {} {\bibfield  {journal} {\bibinfo  {journal} {Nature nanotechnology}\
  }\textbf {\bibinfo {volume} {9}},\ \bibinfo {pages} {257--261} (\bibinfo
  {year} {2014})}\BibitemShut {NoStop}%
\bibitem [{\citenamefont {Ross}\ \emph {et~al.}(2014)\citenamefont {Ross},
  \citenamefont {Klement}, \citenamefont {Jones}, \citenamefont {Ghimire},
  \citenamefont {Yan}, \citenamefont {Mandrus}, \citenamefont {Taniguchi},
  \citenamefont {Watanabe}, \citenamefont {Kitamura}, \citenamefont {Yao},
  \citenamefont {Cobden},\ and\ \citenamefont {Xu}}]{ross2014electrically}%
  \BibitemOpen
  \bibfield  {author} {\bibinfo {author} {\bibfnamefont {J.~S.}\ \bibnamefont
  {Ross}}, \bibinfo {author} {\bibfnamefont {P.}~\bibnamefont {Klement}},
  \bibinfo {author} {\bibfnamefont {A.~M.}\ \bibnamefont {Jones}}, \bibinfo
  {author} {\bibfnamefont {N.~J.}\ \bibnamefont {Ghimire}}, \bibinfo {author}
  {\bibfnamefont {J.}~\bibnamefont {Yan}}, \bibinfo {author} {\bibfnamefont
  {D.~G.}\ \bibnamefont {Mandrus}}, \bibinfo {author} {\bibfnamefont
  {T.}~\bibnamefont {Taniguchi}}, \bibinfo {author} {\bibfnamefont
  {K.}~\bibnamefont {Watanabe}}, \bibinfo {author} {\bibfnamefont
  {K.}~\bibnamefont {Kitamura}}, \bibinfo {author} {\bibfnamefont
  {W.}~\bibnamefont {Yao}}, \bibinfo {author} {\bibfnamefont {D.~H.}\
  \bibnamefont {Cobden}}, \ and\ \bibinfo {author} {\bibfnamefont
  {X.}~\bibnamefont {Xu}},\ }\bibfield  {title} {\enquote {\bibinfo {title}
  {{Electrically tunable excitonic light-emitting diodes based on monolayer
  WSe2 p–n junctions}},}\ }\href {\doibase 10.1038/nnano.2014.26} {\bibfield
  {journal} {\bibinfo  {journal} {Nature Nanotechnology}\ }\textbf {\bibinfo
  {volume} {9}},\ \bibinfo {pages} {268--272} (\bibinfo {year}
  {2014})}\BibitemShut {NoStop}%
\bibitem [{\citenamefont {Novoselov}\ \emph {et~al.}(2004)\citenamefont
  {Novoselov}, \citenamefont {Geim}, \citenamefont {Morozov}, \citenamefont
  {Jiang}, \citenamefont {Zhang}, \citenamefont {Dubonos}, \citenamefont
  {Grigorieva},\ and\ \citenamefont {Firsov}}]{novoselov2004electric}%
  \BibitemOpen
  \bibfield  {author} {\bibinfo {author} {\bibfnamefont {K.~S.}\ \bibnamefont
  {Novoselov}}, \bibinfo {author} {\bibfnamefont {A.~K.}\ \bibnamefont {Geim}},
  \bibinfo {author} {\bibfnamefont {S.~V.}\ \bibnamefont {Morozov}}, \bibinfo
  {author} {\bibfnamefont {D.}~\bibnamefont {Jiang}}, \bibinfo {author}
  {\bibfnamefont {Y.}~\bibnamefont {Zhang}}, \bibinfo {author} {\bibfnamefont
  {S.~V.}\ \bibnamefont {Dubonos}}, \bibinfo {author} {\bibfnamefont {I.~V.}\
  \bibnamefont {Grigorieva}}, \ and\ \bibinfo {author} {\bibfnamefont {A.~A.}\
  \bibnamefont {Firsov}},\ }\bibfield  {title} {\enquote {\bibinfo {title}
  {Electric field effect in atomically thin carbon films},}\ }\href@noop {}
  {\bibfield  {journal} {\bibinfo  {journal} {science}\ }\textbf {\bibinfo
  {volume} {306}},\ \bibinfo {pages} {666--669} (\bibinfo {year}
  {2004})}\BibitemShut {NoStop}%
\bibitem [{\citenamefont {McPherson}\ \emph {et~al.}(2003)\citenamefont
  {McPherson}, \citenamefont {Kim}, \citenamefont {Shanware},\ and\
  \citenamefont {Mogul}}]{mcpherson2003thermochemical}%
  \BibitemOpen
  \bibfield  {author} {\bibinfo {author} {\bibfnamefont {J.}~\bibnamefont
  {McPherson}}, \bibinfo {author} {\bibfnamefont {J.-Y.}\ \bibnamefont {Kim}},
  \bibinfo {author} {\bibfnamefont {A.}~\bibnamefont {Shanware}}, \ and\
  \bibinfo {author} {\bibfnamefont {H.}~\bibnamefont {Mogul}},\ }\bibfield
  {title} {\enquote {\bibinfo {title} {Thermochemical description of dielectric
  breakdown in high dielectric constant materials},}\ }\href@noop {} {\bibfield
   {journal} {\bibinfo  {journal} {Applied Physics Letters}\ }\textbf {\bibinfo
  {volume} {82}},\ \bibinfo {pages} {2121} (\bibinfo {year}
  {2003})}\BibitemShut {NoStop}%
\bibitem [{\citenamefont {Sire}\ \emph {et~al.}(2007)\citenamefont {Sire},
  \citenamefont {Blonkowski}, \citenamefont {Gordon},\ and\ \citenamefont
  {Baron}}]{sire2007statistics}%
  \BibitemOpen
  \bibfield  {author} {\bibinfo {author} {\bibfnamefont {C.}~\bibnamefont
  {Sire}}, \bibinfo {author} {\bibfnamefont {S.}~\bibnamefont {Blonkowski}},
  \bibinfo {author} {\bibfnamefont {M.~J.}\ \bibnamefont {Gordon}}, \ and\
  \bibinfo {author} {\bibfnamefont {T.}~\bibnamefont {Baron}},\ }\bibfield
  {title} {\enquote {\bibinfo {title} {Statistics of electrical breakdown field
  in hfo2 and sio2 films from millimeter to nanometer length scales},}\
  }\href@noop {} {\bibfield  {journal} {\bibinfo  {journal} {Applied Physics
  Letters}\ }\textbf {\bibinfo {volume} {91}},\ \bibinfo {pages} {2905}
  (\bibinfo {year} {2007})}\BibitemShut {NoStop}%
\bibitem [{\citenamefont {Hattori}\ \emph {et~al.}(2014)\citenamefont
  {Hattori}, \citenamefont {Taniguchi}, \citenamefont {Watanabe},\ and\
  \citenamefont {Nagashio}}]{hattori2014layer}%
  \BibitemOpen
  \bibfield  {author} {\bibinfo {author} {\bibfnamefont {Y.}~\bibnamefont
  {Hattori}}, \bibinfo {author} {\bibfnamefont {T.}~\bibnamefont {Taniguchi}},
  \bibinfo {author} {\bibfnamefont {K.}~\bibnamefont {Watanabe}}, \ and\
  \bibinfo {author} {\bibfnamefont {K.}~\bibnamefont {Nagashio}},\ }\bibfield
  {title} {\enquote {\bibinfo {title} {Layer-by-layer dielectric breakdown of
  hexagonal boron nitride},}\ }\href@noop {} {\bibfield  {journal} {\bibinfo
  {journal} {ACS nano}\ }\textbf {\bibinfo {volume} {9}},\ \bibinfo {pages}
  {916--921} (\bibinfo {year} {2014})}\BibitemShut {NoStop}%
\bibitem [{\citenamefont {Efetov}\ and\ \citenamefont
  {Kim}(2010)}]{efetov2010controlling}%
  \BibitemOpen
  \bibfield  {author} {\bibinfo {author} {\bibfnamefont {D.~K.}\ \bibnamefont
  {Efetov}}\ and\ \bibinfo {author} {\bibfnamefont {P.}~\bibnamefont {Kim}},\
  }\bibfield  {title} {\enquote {\bibinfo {title} {Controlling electron-phonon
  interactions in graphene at ultrahigh carrier densities},}\ }\href@noop {}
  {\bibfield  {journal} {\bibinfo  {journal} {Physical review letters}\
  }\textbf {\bibinfo {volume} {105}},\ \bibinfo {pages} {256805} (\bibinfo
  {year} {2010})}\BibitemShut {NoStop}%
\bibitem [{\citenamefont {Ueno}\ \emph {et~al.}(2011)\citenamefont {Ueno},
  \citenamefont {Nakamura}, \citenamefont {Shimotani}, \citenamefont {Yuan},
  \citenamefont {Kimura}, \citenamefont {Nojima}, \citenamefont {Aoki},
  \citenamefont {Iwasa},\ and\ \citenamefont {Kawasaki}}]{ueno2011discovery}%
  \BibitemOpen
  \bibfield  {author} {\bibinfo {author} {\bibfnamefont {K.}~\bibnamefont
  {Ueno}}, \bibinfo {author} {\bibfnamefont {S.}~\bibnamefont {Nakamura}},
  \bibinfo {author} {\bibfnamefont {H.}~\bibnamefont {Shimotani}}, \bibinfo
  {author} {\bibfnamefont {H.}~\bibnamefont {Yuan}}, \bibinfo {author}
  {\bibfnamefont {N.}~\bibnamefont {Kimura}}, \bibinfo {author} {\bibfnamefont
  {T.}~\bibnamefont {Nojima}}, \bibinfo {author} {\bibfnamefont
  {H.}~\bibnamefont {Aoki}}, \bibinfo {author} {\bibfnamefont {Y.}~\bibnamefont
  {Iwasa}}, \ and\ \bibinfo {author} {\bibfnamefont {M.}~\bibnamefont
  {Kawasaki}},\ }\bibfield  {title} {\enquote {\bibinfo {title} {Discovery of
  superconductivity in ktao3 by electrostatic carrier doping},}\ }\href@noop {}
  {\bibfield  {journal} {\bibinfo  {journal} {Nature nanotechnology}\ }\textbf
  {\bibinfo {volume} {6}},\ \bibinfo {pages} {408--412} (\bibinfo {year}
  {2011})}\BibitemShut {NoStop}%
\bibitem [{\citenamefont {Ohno}\ \emph {et~al.}(2009)\citenamefont {Ohno},
  \citenamefont {Maehashi}, \citenamefont {Yamashiro},\ and\ \citenamefont
  {Matsumoto}}]{ohno2009electrolyte}%
  \BibitemOpen
  \bibfield  {author} {\bibinfo {author} {\bibfnamefont {Y.}~\bibnamefont
  {Ohno}}, \bibinfo {author} {\bibfnamefont {K.}~\bibnamefont {Maehashi}},
  \bibinfo {author} {\bibfnamefont {Y.}~\bibnamefont {Yamashiro}}, \ and\
  \bibinfo {author} {\bibfnamefont {K.}~\bibnamefont {Matsumoto}},\ }\bibfield
  {title} {\enquote {\bibinfo {title} {Electrolyte-gated graphene field-effect
  transistors for detecting ph and protein adsorption},}\ }\href@noop {}
  {\bibfield  {journal} {\bibinfo  {journal} {Nano Letters}\ }\textbf {\bibinfo
  {volume} {9}},\ \bibinfo {pages} {3318--3322} (\bibinfo {year}
  {2009})}\BibitemShut {NoStop}%
\bibitem [{\citenamefont {Comsol}(2015)}]{COMSOL}%
  \BibitemOpen
  \bibfield  {author} {\bibinfo {author} {\bibnamefont {Comsol}},\ }\href@noop
  {} {\emph {\bibinfo {title} {COMSOL Multiphysics: Version 5.0}}}\ (\bibinfo
  {publisher} {Comsol},\ \bibinfo {year} {2015})\BibitemShut {NoStop}%
\bibitem [{\citenamefont {Bard}\ \emph {et~al.}(1980)\citenamefont {Bard},
  \citenamefont {Faulkner}, \citenamefont {Leddy},\ and\ \citenamefont
  {Zoski}}]{bard1980electrochemical}%
  \BibitemOpen
  \bibfield  {author} {\bibinfo {author} {\bibfnamefont {A.~J.}\ \bibnamefont
  {Bard}}, \bibinfo {author} {\bibfnamefont {L.~R.}\ \bibnamefont {Faulkner}},
  \bibinfo {author} {\bibfnamefont {J.}~\bibnamefont {Leddy}}, \ and\ \bibinfo
  {author} {\bibfnamefont {C.~G.}\ \bibnamefont {Zoski}},\ }\href@noop {}
  {\emph {\bibinfo {title} {Electrochemical methods: fundamentals and
  applications}}},\ Vol.~\bibinfo {volume} {2}\ (\bibinfo  {publisher} {Wiley
  New York},\ \bibinfo {year} {1980})\BibitemShut {NoStop}%
\bibitem [{\citenamefont {Chen}\ and\ \citenamefont
  {Ahmed}(1993)}]{chen1993fabrication}%
  \BibitemOpen
  \bibfield  {author} {\bibinfo {author} {\bibfnamefont {W.}~\bibnamefont
  {Chen}}\ and\ \bibinfo {author} {\bibfnamefont {H.}~\bibnamefont {Ahmed}},\
  }\bibfield  {title} {\enquote {\bibinfo {title} {Fabrication of 5--7 nm wide
  etched lines in silicon using 100 kev electron-beam lithography and
  polymethylmethacrylate resist},}\ }\href@noop {} {\bibfield  {journal}
  {\bibinfo  {journal} {Applied physics letters}\ }\textbf {\bibinfo {volume}
  {62}},\ \bibinfo {pages} {1499--1501} (\bibinfo {year} {1993})}\BibitemShut
  {NoStop}%
\bibitem [{\citenamefont {Beaumont}\ \emph {et~al.}(1981)\citenamefont
  {Beaumont}, \citenamefont {Bower}, \citenamefont {Tamamura},\ and\
  \citenamefont {Wilkinson}}]{beaumont1981sub}%
  \BibitemOpen
  \bibfield  {author} {\bibinfo {author} {\bibfnamefont {S.}~\bibnamefont
  {Beaumont}}, \bibinfo {author} {\bibfnamefont {P.}~\bibnamefont {Bower}},
  \bibinfo {author} {\bibfnamefont {T.}~\bibnamefont {Tamamura}}, \ and\
  \bibinfo {author} {\bibfnamefont {C.}~\bibnamefont {Wilkinson}},\ }\bibfield
  {title} {\enquote {\bibinfo {title} {Sub-20-nm-wide metal lines by
  electron-beam exposure of thin poly (methyl methacrylate) films and
  liftoff},}\ }\href@noop {} {\bibfield  {journal} {\bibinfo  {journal}
  {Applied Physics Letters}\ }\textbf {\bibinfo {volume} {38}},\ \bibinfo
  {pages} {436--439} (\bibinfo {year} {1981})}\BibitemShut {NoStop}%
\bibitem [{\citenamefont {Duan}\ \emph {et~al.}(2010)\citenamefont {Duan},
  \citenamefont {Winston}, \citenamefont {Yang}, \citenamefont {Cord},
  \citenamefont {Manfrinato},\ and\ \citenamefont {Berggren}}]{duan2010sub}%
  \BibitemOpen
  \bibfield  {author} {\bibinfo {author} {\bibfnamefont {H.}~\bibnamefont
  {Duan}}, \bibinfo {author} {\bibfnamefont {D.}~\bibnamefont {Winston}},
  \bibinfo {author} {\bibfnamefont {J.~K.}\ \bibnamefont {Yang}}, \bibinfo
  {author} {\bibfnamefont {B.~M.}\ \bibnamefont {Cord}}, \bibinfo {author}
  {\bibfnamefont {V.~R.}\ \bibnamefont {Manfrinato}}, \ and\ \bibinfo {author}
  {\bibfnamefont {K.~K.}\ \bibnamefont {Berggren}},\ }\bibfield  {title}
  {\enquote {\bibinfo {title} {Sub-10-nm half-pitch electron-beam lithography
  by using poly (methyl methacrylate) as a negative resist},}\ }\href@noop {}
  {\bibfield  {journal} {\bibinfo  {journal} {Journal of Vacuum Science \&
  Technology B}\ }\textbf {\bibinfo {volume} {28}},\ \bibinfo {pages}
  {C6C58--C6C62} (\bibinfo {year} {2010})}\BibitemShut {NoStop}%
\bibitem [{\citenamefont {Duan}, \citenamefont {Xie},\ and\ \citenamefont
  {Han}(2008)}]{duan2008turning}%
  \BibitemOpen
  \bibfield  {author} {\bibinfo {author} {\bibfnamefont {H.}~\bibnamefont
  {Duan}}, \bibinfo {author} {\bibfnamefont {E.}~\bibnamefont {Xie}}, \ and\
  \bibinfo {author} {\bibfnamefont {L.}~\bibnamefont {Han}},\ }\bibfield
  {title} {\enquote {\bibinfo {title} {Turning electrospun poly (methyl
  methacrylate) nanofibers into graphitic nanostructures by in situ electron
  beam irradiation},}\ }\href@noop {} {\bibfield  {journal} {\bibinfo
  {journal} {Journal of Applied Physics}\ }\textbf {\bibinfo {volume} {103}},\
  \bibinfo {pages} {046105} (\bibinfo {year} {2008})}\BibitemShut {NoStop}%
\bibitem [{\citenamefont {Efetov}(2014)}]{efetov2014towards}%
  \BibitemOpen
  \bibfield  {author} {\bibinfo {author} {\bibfnamefont {D.~K.}\ \bibnamefont
  {Efetov}},\ }\bibfield  {title} {\enquote {\bibinfo {title} {Towards inducing
  superconductivity into graphene},}\ }\href@noop {} {\  (\bibinfo {year}
  {2014})}\BibitemShut {NoStop}%
\bibitem [{\citenamefont {Williams}, \citenamefont {DiCarlo},\ and\
  \citenamefont {Marcus}(2007)}]{williams2007quantum}%
  \BibitemOpen
  \bibfield  {author} {\bibinfo {author} {\bibfnamefont {J.}~\bibnamefont
  {Williams}}, \bibinfo {author} {\bibfnamefont {L.}~\bibnamefont {DiCarlo}}, \
  and\ \bibinfo {author} {\bibfnamefont {C.}~\bibnamefont {Marcus}},\
  }\bibfield  {title} {\enquote {\bibinfo {title} {Quantum hall effect in a
  gate-controlled pn junction of graphene},}\ }\href@noop {} {\bibfield
  {journal} {\bibinfo  {journal} {Science}\ }\textbf {\bibinfo {volume}
  {317}},\ \bibinfo {pages} {638--641} (\bibinfo {year} {2007})}\BibitemShut
  {NoStop}%
\bibitem [{\citenamefont {Song}\ \emph {et~al.}(2011)\citenamefont {Song},
  \citenamefont {Rudner}, \citenamefont {Marcus},\ and\ \citenamefont
  {Levitov}}]{song2011hot}%
  \BibitemOpen
  \bibfield  {author} {\bibinfo {author} {\bibfnamefont {J.~C.}\ \bibnamefont
  {Song}}, \bibinfo {author} {\bibfnamefont {M.~S.}\ \bibnamefont {Rudner}},
  \bibinfo {author} {\bibfnamefont {C.~M.}\ \bibnamefont {Marcus}}, \ and\
  \bibinfo {author} {\bibfnamefont {L.~S.}\ \bibnamefont {Levitov}},\
  }\bibfield  {title} {\enquote {\bibinfo {title} {Hot carrier transport and
  photocurrent response in graphene},}\ }\href@noop {} {\bibfield  {journal}
  {\bibinfo  {journal} {Nano letters}\ }\textbf {\bibinfo {volume} {11}},\
  \bibinfo {pages} {4688--4692} (\bibinfo {year} {2011})}\BibitemShut {NoStop}%
\bibitem [{\citenamefont {Hsu}\ \emph {et~al.}(2015)\citenamefont {Hsu},
  \citenamefont {Herring}, \citenamefont {Gabor}, \citenamefont {Ha},
  \citenamefont {Shin}, \citenamefont {Song}, \citenamefont {Chin},
  \citenamefont {Dubey}, \citenamefont {Chandrakasan}, \citenamefont {Kong},
  \citenamefont {Jarillo-Herrero},\ and\ \citenamefont
  {Palacios}}]{hsu2015graphene}%
  \BibitemOpen
  \bibfield  {author} {\bibinfo {author} {\bibfnamefont {A.~L.}\ \bibnamefont
  {Hsu}}, \bibinfo {author} {\bibfnamefont {P.}~\bibnamefont {Herring}},
  \bibinfo {author} {\bibfnamefont {N.}~\bibnamefont {Gabor}}, \bibinfo
  {author} {\bibfnamefont {S.}~\bibnamefont {Ha}}, \bibinfo {author}
  {\bibfnamefont {Y.~C.}\ \bibnamefont {Shin}}, \bibinfo {author}
  {\bibfnamefont {Y.}~\bibnamefont {Song}}, \bibinfo {author} {\bibfnamefont
  {M.}~\bibnamefont {Chin}}, \bibinfo {author} {\bibfnamefont {M.}~\bibnamefont
  {Dubey}}, \bibinfo {author} {\bibfnamefont {A.}~\bibnamefont {Chandrakasan}},
  \bibinfo {author} {\bibfnamefont {J.}~\bibnamefont {Kong}}, \bibinfo {author}
  {\bibfnamefont {P.}~\bibnamefont {Jarillo-Herrero}}, \ and\ \bibinfo {author}
  {\bibfnamefont {T.}~\bibnamefont {Palacios}},\ }\bibfield  {title} {\enquote
  {\bibinfo {title} {Graphene-based thermopile for thermal imaging
  applications},}\ }\href@noop {} {\bibfield  {journal} {\bibinfo  {journal}
  {Nano letters}\ }\textbf {\bibinfo {volume} {15}},\ \bibinfo {pages}
  {7211--7216} (\bibinfo {year} {2015})}\BibitemShut {NoStop}%
\bibitem [{\citenamefont {Herring}\ \emph {et~al.}(2014)\citenamefont
  {Herring}, \citenamefont {Hsu}, \citenamefont {Gabor}, \citenamefont {Shin},
  \citenamefont {Kong}, \citenamefont {Palacios},\ and\ \citenamefont
  {Jarillo-Herrero}}]{herring2014photoresponse}%
  \BibitemOpen
  \bibfield  {author} {\bibinfo {author} {\bibfnamefont {P.~K.}\ \bibnamefont
  {Herring}}, \bibinfo {author} {\bibfnamefont {A.~L.}\ \bibnamefont {Hsu}},
  \bibinfo {author} {\bibfnamefont {N.~M.}\ \bibnamefont {Gabor}}, \bibinfo
  {author} {\bibfnamefont {Y.~C.}\ \bibnamefont {Shin}}, \bibinfo {author}
  {\bibfnamefont {J.}~\bibnamefont {Kong}}, \bibinfo {author} {\bibfnamefont
  {T.}~\bibnamefont {Palacios}}, \ and\ \bibinfo {author} {\bibfnamefont
  {P.}~\bibnamefont {Jarillo-Herrero}},\ }\bibfield  {title} {\enquote
  {\bibinfo {title} {Photoresponse of an electrically tunable ambipolar
  graphene infrared thermocouple},}\ }\href@noop {} {\bibfield  {journal}
  {\bibinfo  {journal} {Nano letters}\ }\textbf {\bibinfo {volume} {14}},\
  \bibinfo {pages} {901--907} (\bibinfo {year} {2014})}\BibitemShut {NoStop}%
\bibitem [{\citenamefont {Henriksen}, \citenamefont {Nandi},\ and\
  \citenamefont {Eisenstein}(2012)}]{henriksen2012quantum}%
  \BibitemOpen
  \bibfield  {author} {\bibinfo {author} {\bibfnamefont {E.}~\bibnamefont
  {Henriksen}}, \bibinfo {author} {\bibfnamefont {D.}~\bibnamefont {Nandi}}, \
  and\ \bibinfo {author} {\bibfnamefont {J.}~\bibnamefont {Eisenstein}},\
  }\bibfield  {title} {\enquote {\bibinfo {title} {Quantum hall effect and
  semimetallic behavior of dual-gated aba-stacked trilayer graphene},}\
  }\href@noop {} {\bibfield  {journal} {\bibinfo  {journal} {Physical Review
  X}\ }\textbf {\bibinfo {volume} {2}},\ \bibinfo {pages} {011004} (\bibinfo
  {year} {2012})}\BibitemShut {NoStop}%
\bibitem [{\citenamefont {Low}\ and\ \citenamefont
  {Avouris}(2014)}]{low2014graphene}%
  \BibitemOpen
  \bibfield  {author} {\bibinfo {author} {\bibfnamefont {T.}~\bibnamefont
  {Low}}\ and\ \bibinfo {author} {\bibfnamefont {P.}~\bibnamefont {Avouris}},\
  }\bibfield  {title} {\enquote {\bibinfo {title} {Graphene plasmonics for
  terahertz to mid-infrared applications},}\ }\href@noop {} {\bibfield
  {journal} {\bibinfo  {journal} {ACS nano}\ }\textbf {\bibinfo {volume} {8}},\
  \bibinfo {pages} {1086--1101} (\bibinfo {year} {2014})}\BibitemShut {NoStop}%
\end{thebibliography}%

\end{document}